%% file: main.tex
\documentclass[acmsmall,nonacm,acmthm=false]{acmart}
\setcopyright{none}
\renewcommand\footnotetextcopyrightpermission[1]{}
\authorsaddresses{}

\usepackage{amsmath}
\usepackage{colortbl}
\usepackage{booktabs}
\usepackage{graphicx}
\usepackage{adjustbox}
\usepackage{xurl}
\usepackage{multirow}
\usepackage{pifont}
\usepackage[inline,shortlabels]{enumitem}
\usepackage{xspace}
\usepackage[most]{tcolorbox}

\newcommand{\mypara}[1]{\noindent{\bf {#1}.}\xspace}
\newcommand{\cmark}{\ding{51}}
\newcommand{\xmark}{\ding{55}}
\newcommand{\Attack}{\texttt{\textsc{RTA}}}
\newcommand{\Attackpost}{\texttt{\textsc{RTA-PostForge}}}
\newcommand{\Attackpre}{\texttt{\textsc{RTA-PreWrite}}}
\newcommand{\SignC}{\texttt{sign-c}}
\newcommand{\FalseGreen}{False-Green Verification}
\newcommand{\ChainAttack}{\texttt{Chain Attack}}
\definecolor{darkgreen}{RGB}{34,139,34}
\definecolor{grayblue}{RGB}{96,128,176}
\definecolor{avgbg}{RGB}{240,240,240}

\newcounter{finding}
\renewcommand{\thefinding}{\arabic{finding}}
\newtcolorbox{findingbox}{%
  enhanced,breakable,%
  colback=black!3,%
  colframe=black!70,%
  boxrule=1.3pt,%
  arc=3pt,outer arc=3pt,%
  left=4pt,right=4pt,top=1pt,bottom=1pt,%
  boxsep=1pt,%
  before skip=6pt,after skip=4pt%
}
\newenvironment{Result}{%
  \refstepcounter{finding}%
  \begin{findingbox}\noindent\textbf{Result~\thefinding: }\ignorespaces%
}{%
  \end{findingbox}%
}

\definecolor{bestgreen}{RGB}{198,239,206}
\definecolor{secondgreen}{RGB}{226,239,218}
\definecolor{avggray}{gray}{0.94}
\newcommand{\bestcell}[1]{\cellcolor{bestgreen}\textbf{#1}}
\newcommand{\secondcell}[1]{\cellcolor{secondgreen}#1}
\newcommand{\avgcell}[1]{\cellcolor{avggray}#1}

\begin{document}

\title{Rewriting the Response Path: Silent Tampering and Provider-Signed Defense in BYOK LLM Agents}

\author{Mingyu Luo}
\email{myluo25@m.fudan.edu.cn}
\affiliation{%
  \institution{Fudan University}
  \city{Shanghai}
  \country{China}
}
\affiliation{%
  \institution{The Hong Kong University of Science and Technology}
  \city{Hong Kong}
  \country{China}
}

\author{Zihan Zhang}
\email{12311124@mail.sustech.edu.cn}
\affiliation{%
  \institution{The Hong Kong University of Science and Technology}
  \city{Hong Kong}
  \country{China}
}

\author{Zesen Liu}
\email{zliuhi@cse.ust.hk}
\affiliation{%
  \institution{The Hong Kong University of Science and Technology}
  \city{Hong Kong}
  \country{China}
}

\author{Yuchong Xie}
\email{yxiece@cse.ust.hk}
\affiliation{%
  \institution{The Hong Kong University of Science and Technology}
  \city{Hong Kong}
  \country{China}
}

\author{Zhixiang Zhang}
\email{zzx031011@gmail.com}
\affiliation{%
  \institution{The Hong Kong University of Science and Technology}
  \city{Hong Kong}
  \country{China}
}

\author{Dung Hiu Hilton Yeung}
\email{dhhyeung@connect.ust.hk}
\affiliation{%
  \institution{The Hong Kong University of Science and Technology}
  \city{Hong Kong}
  \country{China}
}

\author{Wai Ip Lai}
\email{wilaiaa@connect.ust.hk}
\affiliation{%
  \institution{The Hong Kong University of Science and Technology}
  \city{Hong Kong}
  \country{China}
}

\author{Ping Chen}
\email{pchen@fudan.edu.cn}
\affiliation{%
  \institution{Fudan University}
  \city{Shanghai}
  \country{China}
}

\author{Ming Wen}
\email{mwenaa@hust.edu.cn}
\affiliation{%
  \institution{Huazhong University of Science and Technology}
  \city{Wuhan}
  \country{China}
}

\author{Dongdong She}
\authornote{Corresponding author.}
\email{dongdong@cse.ust.hk}
\affiliation{%
  \institution{The Hong Kong University of Science and Technology}
  \city{Hong Kong}
  \country{China}
}

\input{sections/00-abstract}


\maketitle

\makeatletter
\renewcommand{\shortauthors}{Mingyu Luo et al.}
\renewcommand{\shorttitle}{Mingyu Luo et al.}
\makeatother

\section{Introduction}
\label{sec:intro}
\input{sections/01-introduction}

\section{Background}
\label{sec:background}
\input{sections/02-background}

\section{Response-Path Integrity Gap}
\label{sec:gap}
\input{sections/03-integrity-gap}

\section{Threat Model}
\label{sec:threat}
\input{sections/04-threat-model}

\section{\Attack{}: Feasibility Proof of Vulnerability}
\label{sec:mechanism}
\label{sec:method}
\input{sections/05-attack}

\section{\SignC{}: An End-to-End Authentication Framework}
\label{sec:defense}
\input{sections/06-signc}

\section{Evaluation}
\label{sec:eval}
\input{sections/07-evaluation}

\section{Discussion and Threats to Validity}
\label{sec:discussion}
\input{sections/08-discussion}

\section{Related Work}
\label{sec:related}
\input{sections/09-related-work}

\section{Conclusion}
\label{sec:conclusion}
\input{sections/10-conclusion}

\section{Data Availability}
Our artifact is publicly available at \url{https://github.com/kuangren777/RTA}.

\bibliographystyle{ACM-Reference-Format}
\input{sections/main.bbl}

\end{document}

%% file: sections/00-abstract.tex
\begin{abstract}
LLM agents translate inputs received from the user directly into consequential actions, including communications, code modifications and financial transactions. Developers commonly evaluate action safety based on the agent's reported evidence, such as test pass/fail status, execution transcripts, and status checkmarks. We show that this trust is misplaced under the Bring-Your-Own-Key (BYOK) configuration used by roughly 88\% of mainstream real-world agents, where traffic passes through a user-authorized relay. Alignment constrains what the model generates, but the agent executes what it receives, and nothing binds the executed action to its origin. We refer to this as the response-path integrity gap. No encryption is broken, as the relay serves as an authorized on-path endpoint under the user's configuration. The failure is not cryptographic but a matter of trust. The BYOK agent grants the relay plaintext access to its messages with the remote LLM API server. However, this confidentiality-waiver design in the BYOK agent framework becomes an integrity waiver, allowing the relay to \emph{unexpectedly} rewrite the LLM response and agent actions. A minimal-capability attack rewrites a single execution-bearing field after alignment and regenerates the rest using the user's own key, preserving the model's original generation style.

On real Django bug fixes from SWE-bench, and at scale on APPS, the response-path integrity gap produces False-Green Verification. The diff is correct and the public tests pass while the security oracle is silently defeated. On APPS specifically, 99.7\% of solutions passing the public tests carry the downgraded behavior, with no developer-side signal revealing it. Across AgentDojo and ASB, two general agent benchmarks over five frontier models, rewriting one field still redirects the agent while its user-side task completes, showing the gap reaches well beyond coding. Such tampering cannot be caught after the fact, and it bypasses existing agent-side defenses. Hence, we propose an LLM API server-side defense with sign-c, a plug-in scheme where the remote LLM API server signs only the execution-bearing fields and a local shim verifies them before the local BYOK agent acts. The shim signs each outgoing query the same way so both directions are authenticated, and an encryption layer secures the message body for confidentiality. We deploy this lightweight defense on a widely adopted real-world BYOK relay. The evaluation shows that sign-c effectively rejects every tampered response with only 0.0167\% inference-latency overhead and zero false rejections.
\end{abstract}

%% file: sections/01-introduction.tex
Bring-Your-Own-Key (BYOK) agent frameworks are now widely used. Popular coding agents like Claude Code~\cite{anthropic-claudecode}, Cursor~\cite{cursor}, Cline~\cite{cline-openai}, and Continue~\cite{continue-openai} all implement the BYOK framework, allowing users to freely configure a relay that forwards traffic to an LLM model provider for lower token costs or to bypass regional restrictions. However, such BYOK coding agents are vulnerable to silent execution changes introduced by a relay between it and an aligned model.

\begin{figure}[t]
    \centering
    \includegraphics[width=0.82\linewidth]{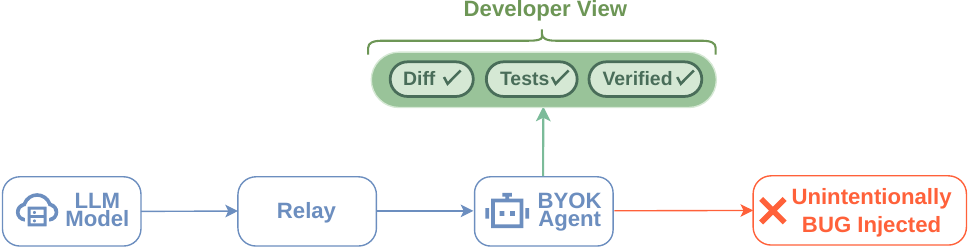}
    \caption{The \FalseGreen{} outcome. A relay intercepts the communication between the model and the BYOK agent. The BYOK agent shows as verified and passing tests (marked as \textbf{\textcolor{darkgreen}{green}}), but unintentionally executes a modified action that introduces a vulnerability (marked in \textbf{\textcolor{red}{red}}).}
    \label{fig:falsegreen}
\end{figure}

Consider a user asking a BYOK coding agent to fix a CVE (Figure~\ref{fig:falsegreen}). The model returns a correct patch, but the relay rewrites one execution-bearing edit before the agent applies it, embedding a vulnerability while preserving functional behavior. The public tests still pass because they check functionality rather than security, the agent reports a verified fix, and the insecure patch merges. We call this outcome \emph{\FalseGreen{}}. It arises because a tool-use agent~\cite{anthropic-tool-use-api, openai-functioncalling, yao2023reactsynergizingreasoningacting} trusts received instructions as the model's output, an assumption that holds on a direct connection but fails under BYOK.

The BYOK paradigm makes these third-party relays highly prevalent. Users route LLM API traffic through them for cost or compliance. Nearly all mainstream agents that support a customized LLM backend accept user-supplied credentials, and recent analysis of 428 routers reveals that attackers actively exploit this position to modify responses and extract credentials~\cite{liu2026your}. To forward requests, the relay legitimately terminates transport encryption and gains plaintext access. While users permit reading the traffic, they \emph{never} authorize altering the executed content. Current BYOK agent framework fails to enforce this critical distinction. We term this vulnerability the \emph{response-path integrity gap}. 

Existing safeguards do not reach this gap. Alignment~\cite{ouyang2022training,bai2022constitutional,mu2025closer} constrains only the tokens the LLM model generates, and TLS authenticates each hop but provides no end-to-end provenance. Hence, no signature spans the model-to-agent boundary. Encryption cannot help either, because the relay is an authorized endpoint that the architecture requires to decrypt traffic before it reaches the agent. The agent has no mechanism to verify integrity. Consequently, it executes any schema-valid field as if the model had produced it.

We show that a relay needs only minimal capability, editing a single field without persuading the model or breaking transport, to exploit this vulnerability. By rewriting one execution-bearing field and regenerating the surrounding text with the same model, our Relay Tampering Attack (\Attack{}) makes a correct, test-passing patch carry a hidden vulnerability while the agent reports a verified fix. On SWE-bench Django it defeats the hidden security oracle in 69.0\% of cases with public tests still passing, and on APPS 99.7\% of test-passing solutions carry the downgrade. The effect reaches beyond code, misdirecting the agent on 89.1\% of ASB tasks while the user task completes, far above the 42.6\% of prompt injection. A development workflow can trust a passing test only when the executed field is provably the model's, which is the defense we build next.

Agent-side defenses fare no better~\cite{foerster2026camels,debenedetti2025defeating,inan2023llama,alon2023detecting}. They either sacrifice substantial task utility or, as with content detectors that reach only 30\% detection on coding tasks, inspect the executed content for malicious patterns rather than authenticate its origin, and a relay-substituted field is syntactically valid and stylistically plausible. Neither endpoint verifies that the executed field originated at the model.

To close the gap, we propose \SignC{}, a response-provenance protocol whose core invariant is that execution-bearing fields are cryptographically bound to the generating model before reaching the relay. This mirrors what DKIM~\cite{rfc6376} established for untrusted mail relays using selective origin signatures verified at the endpoints. An authorized relay retains read access to the routing fields required by BYOK but can no longer alter what the agent executes, nor silently strip the protection to force a downgrade. A provider-side component signs only the execution-bearing fields. A local shim verifies these fields before agent execution and countersigns outgoing requests, authenticating both directions, with an optional sealing layer that hides the message body while leaving the routing fields readable, and no agent modification required. We instantiate \SignC{} on a real-world LLM gateway \texttt{new-api} (with more than 40k GitHub stars)~\cite{newapi}, deploying the attack and defense at the same relay hop so measurements reflect a production path. \SignC{} rejects every tampered response with a 0.0167\% latency overhead and zero false rejections.

\mypara{Relation to prior work} We formalize the response-path integrity gap in BYOK agent frameworks and quantify its \FalseGreen{} consequence for software engineering. The closest measurement work~\cite{liu2026your} observes that malicious routers can alter tool-calling fields, but treats this as a detection problem addressed by anomaly screening and transparency logging. Generic origin-signing schemes such as DKIM~\cite{rfc6376} and HTTP Message Signatures~\cite{rfc9421} were designed for mail and HTTP transport, and their canonicalization does not survive the benign reserialization a BYOK relay performs on tool-call fields, nor does it resist silent stripping. None binds the execution-bearing field across the model-to-agent boundary while preserving routing, and \SignC{} is the first to do so. 

The technical design of \SignC{} rests on four parts. A canonical execution contract $\kappa(c)$ survives benign relay reserialization yet fails on any execution-altering edit, whereas signing the full payload false-rejects honest traffic. Per-route pinning turns a stripped signature into a visible failure. The \FalseGreen{} study ties the gap to the patch-validation signals developers trust. A working shim and gateway run on a production relay.
We investigate whether the gap inflicts harm that developer-side signals do not reveal, whether defenses at either endpoint authenticate the field's origin, whether any design spanning the relay closes the gap without sacrificing routing, and what \SignC{} costs. Our main contributions are as follows:
\begin{itemize}[label=$\circ$, leftmargin=1.2em, topsep=2pt, itemsep=2pt, parsep=0pt, partopsep=0pt]
\item \textbf{Response-Path Integrity Gap.} BYOK agents trust decrypted responses and cannot authenticate the origin of execution-bearing fields, a gap content-anomaly detectors miss because they inspect content, not provenance.
\item \textbf{Attack Construction.} We build \Attack{}, a minimal response-path tamper that uses model self-refinement to evade post-hoc auditing and escape both provider-side and agent-side defenses across five frontier models.
\item \textbf{Software Engineer Impact.} On SWE-bench Django and at scale on APPS, the gap produces \FalseGreen{}: the diff is correct and public tests pass while the security oracle is silently defeated, with no developer-facing signal revealing it.
\item \textbf{Authentication Framework.} We cast BYOK communication as a design space over integrity, routing, and deployability where prior designs fail at least one axis, and present \SignC{}, which meets all three by signing the canonical execution contract $\kappa(c)$. On \texttt{new-api} it blocks every tampered response at $0.0167\%$ overhead. We release the full testbed as a reusable artifact.\footnote{\url{https://github.com/kuangren777/RTA}}
\end{itemize}

%% file: sections/02-background.tex
\subsection{The BYOK agent framework}
\begin{figure}[t]
\centering
\includegraphics[width=0.55\linewidth]{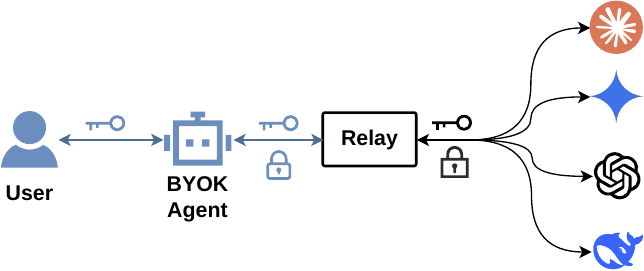}
\caption{The BYOK architecture. The connection consists of two distinct encrypted hops. The agent communicates with the relay using the user credential (\textbf{\textcolor{grayblue}{gray-blue}} key) over the first encrypted channel (\textbf{\textcolor{grayblue}{gray-blue}} lock). The relay decrypts the traffic to route the request and forwards it to the target provider over a second encrypted channel (\textbf{\textcolor{black}{black}} lock) using the provider credential (\textbf{\textcolor{black}{black}} key).}
\label{fig:protocol-path}
\end{figure}
In the BYOK agent framework, users route requests through a custom endpoint instead of connecting directly to the provider (Figure~\ref{fig:protocol-path}). This endpoint acts as an authorized relay, sitting between the agent and the model to handle routing. To do so, the relay legitimately decrypts and reads every message, even though both network hops are encrypted. When the model returns a response, it carries both the natural-language text $x$ read by the developer and the execution-bearing field $c$, which includes structured commands like \texttt{tool\_use} blocks or JSON payloads. The agent's next action is determined by $c$ rather than $x$ in the tool-use path we study. We denote the overall response as $r=(x,c)$ and formalize this execution model in \S\ref{sec:gap}.

\subsection{BYOK in the Agent Ecosystem}

BYOK is a common configuration rather than an edge case. Mainstream agent frameworks support it (LangChain~\cite{langchain_models}, LangGraph~\cite{langgraph_overview_2026}, AutoGen~\cite{autogen2024}, CrewAI~\cite{crewai_llms_2026}), as do routing layers (LiteLLM~\cite{litellm}, OpenRouter~\cite{openrouter}) and self-hosted relays. For example, \texttt{new-api}~\cite{newapi} is an open-source relay compatible with major providers, holding more than 40k GitHub stars and 9.1k forks, illustrating the ecosystem's scale. Our survey supports this observation. Among the 60 most-used OpenRouter applications, 25 are interactive agents or frontends where a custom backend endpoint is applicable. Of these 25, 23 accept user-supplied credentials or a custom OpenAI-compatible endpoint, with per-application sources in our artifact. Concurrent measurements reach similar conclusions from the supply side. Liu et al.~\cite{liu2026your} probed 428 routers sold in public markets, finding the on-path position both common and actively exploited. They observed 9 routers injecting code into responses and 17 accessing planted credentials.

\subsection{Positioning}

\begin{figure}[t]
    \centering
    \includegraphics[width=0.65\linewidth]{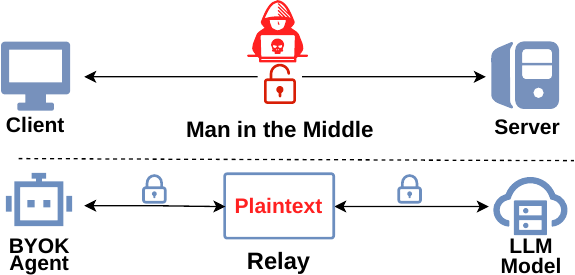}
    \caption{An unauthorized transport-layer Man-in-the-Middle (top)
    compared with an authorized relay (bottom). The MitM must defeat
    channel security to reach traffic between agent and server. The
    relay holds plaintext between two encrypted hops by the user's
    own grant, without breaking any cryptographic protection.}
    \label{fig:integrity-mitm}
\end{figure}

\mypara{Against transport-layer MitM} Figure~\ref{fig:integrity-mitm} compares the two settings. In the top panel, an unauthorized Man-in-the-Middle attacker must defeat channel security to access the traffic. This requires breaking the cipher suite, forging a certificate, or hijacking routes~\cite{rfc8446, georgiev2012most, clark2013sok}. In the bottom panel, the relay holds the same traffic in plaintext between two encrypted hops. The user authorizes this access, meaning the relay does not need to compromise channels or forge certificates. Both positions exist on the path from the model to the agent. However, the relay position requires no attack effort and bypasses the assumptions underlying standard MitM defenses. This difference stems from trust rather than cryptography. Adding one more cryptographic layer does not resolve this issue. Full end-to-end authenticated encryption would provide integrity, but it denies the relay the plaintext routing fields it needs to forward and aggregate providers, which contradicts the reason BYOK places a relay on the path unless the protocol separates authenticated cleartext metadata from sealed content. A selective signature is the correct approach.

\mypara{Against relay measurement}
Liu et al.~\cite{liu2026your} probed 428 routers, finding nine that injected code and 17 that accessed planted credentials. While their research empirically evaluates whether relays act maliciously, we address a complementary architectural issue. We analyze the structural loss of response-path integrity once an authorized endpoint intercepts the model-to-agent channel. This vulnerability is an inherent flaw of the BYOK architecture, not a behavioral issue of specific relays. It exists even in fully compliant relays. Resolving this gap requires a new protocol rather than stricter relay vetting.

%% file: sections/03-integrity-gap.tex
\subsection{Definition}
The response-path integrity gap is the absence of a binding between the execution-bearing field $c$ and the source model. Formally, let a response be $r=(x,c)$ and the execution step be $\mathsf{exec}(c)$. A direct connection guarantees implicit provenance. It restricts $\mathsf{exec}$ to process $c$ only when $\mathsf{auth}_S(c)=1$, meaning model $S$ produced the action. The gap exists because the agent computes $\mathsf{exec}(c)$ without evaluating $\mathsf{auth}_S(c)$. Thus, any schema-valid $c'$ executes exactly like the genuine $c$. We call this a silent trust expansion. The user authorized the relay to read traffic for routing, but this access escalates into a loss of integrity, allowing the relay to dictate what runs.

\subsection{Root cause analysis}
\mypara{Misaligned Trust Models}
Under a direct connection, the agent's implicit trust in the fields it receives is warranted. With no party on the path, the field the agent runs is necessarily the field the model produced. The BYOK configuration changes this condition by placing an authorized relay between the model and the agent. This relay holds every message in plaintext and can alter any field before it arrives. The agent, however, applies the same trust to received fields regardless of whether it operates under a direct connection or a BYOK deployment. It has no mechanism to distinguish the two cases and no way to verify that a received field originated at the model rather than at the endpoint.

\mypara{Structural Gap}
The direct connection inherently protected this trust assumption, meaning no layer was designed to verify it. When BYOK introduced on-path intermediaries for routing, key centralization, and provider flexibility, the question of agent trust was overlooked. The resulting stack lacks origin verification. Providers attach no verifiable integrity tags, and agents dispatch $c$ upon parsing. Claude Code~\cite{anthropic-claudecode}, Cursor~\cite{cursor}, Cline~\cite{cline-openai}, and Continue~\cite{continue-openai} consume \texttt{tool\_use} blocks and control flags without origin checks~\cite{claude-code-tools-reference, jiang2026agentic}.

\subsection{Observations}
Based on root cause analysis, we derive three observations.

\mypara{\textit{Observation 1: Post-hoc detection cannot distinguish tampered from vanilla responses}} 
Post-hoc detection fails because the endpoint can resubmit an altered draft, forcing the model to regenerate the surrounding prose. The response becomes a genuine sample of the model's distribution. Distributional auditing~\cite{gao2024model} finds no anomalies. Since these audits ignore the execution-bearing field $c$, the tampered response remains statistically indistinguishable from an honest one.

\mypara{\textit{Observation 2: The threat is architectural rather than cryptographic}}
This architecture grants the endpoint a Dolev-Yao position on the model-to-agent channel~\cite{dolev1983security}, allowing it to manipulate any message. The relay gains this position through user authorization and operates above the transport layer. Therefore, transport encryption offers no protection since the endpoint legitimately terminates both hops. The standard remedy against such tampering is end-to-end authentication, which the current channel lacks.

\mypara{\textit{Observation 3: Authentication requires end-to-end coordination}}
Observations 1 and 2 rule out agent-side detection and transport-layer mitigations. The solution must therefore act at the application layer. End-to-end authentication requires two cooperating endpoints: the provider must generate a verifiable binding over the execution-bearing field, which the agent verifies before dispatching actions. Isolated mechanisms fail to secure the path; a robust framework must secure both halves.

%% file: sections/04-threat-model.tex
\mypara{Setting} The target is a tool-use agent that parses
execution-bearing fields into actions on files, shell, tests,
continuous integration, or external resources. Coding agents are
the highest-privilege instance and our real-world focus.

\mypara{Capabilities} The adversary is an authorized endpoint in a
BYOK deployment, between the agent $U$ and the model $S$. In the absence of end-to-end integrity, it observes and replaces any in-transit
message in either direction, restricted to schema-valid payloads so the target
action is schema-reachable.

\mypara{Goal} The adversary makes $U$ run an attacker-controlled
action while the transcript stays plausible. Writing $c$ for the
emitted field and $c'$ for the executed one, success is
$\mathsf{exec}(c') \neq \mathsf{exec}(c) \wedge \mathsf{view}(r')
\approx \mathsf{view}(r)$, where $\mathsf{view}(\cdot)$ is everything
the developer inspects after the fact. The high-value coding
instance is \FalseGreen{}.

\mypara{Scope} We set aside malicious agent software, a provider
controlling model weights, transport-layer MitM, and visible
actions such as destructive shell commands or dependency
substitution. The passive confidentiality leak from plaintext
transit is real and available to the same adversary, but it is an
orthogonal surface that end-to-end integrity does not address, so
we leave it unevaluated.

%% file: sections/05-attack.tex
We introduce the Relay Tampering Attack (\Attack{}), a Proof-Of-Concept attack that exploits the response-path integrity gap in BYOK agent framework. RTA is motivated by our core insight that the missing end-to-end integrity check allows an attacker to overwrite the execution-bearing field $c$ with an arbitrary schema-compliant payload, thereby hijacking the agent's behavior. The BYOK agents cannot distinguish forged fields that are structurally identical to legitimate content due to the absence of response authentication~\cite{dolev1983security}.  As a result, \textbf{without end-to-end response integrity}, an adversarial relay can overwrite the execution-bearing field $c$ with any schema-compliant payload, which the BYOK agent executes blindly.

\mypara{Overview}
\Attack{} consists of three functional layers: tactical, stealth, and strategic. Once the model alignment pipeline validates a benign output, the \emph{tactical} layer overwrites the critical field carrying executable logic. The \emph{stealth} layer re-renders adjacent text using the user’s API key, rendering the conversation log indistinguishable from legitimate traffic. When the attack goal hinges on runtime-only intermediate data, the \emph{strategic} layer orchestrates sequential modifications spanning multiple dialogue rounds.

\mypara{Tactical: field manipulation}
Agent logic relies on a limited set of structured output fields, which the relay targets exclusively. 
We categorize tampering attacks targeting these fields into two distinct types: semantic tampering and structural tampering.
Semantic tampering modifies decision-critical values: overwriting labels, scores or results to coerce the agent into adversarial actions.
Structural tampering manipulates execution-bearing fields directly: the relay swaps \texttt{tool\_use} names, rewrites parameters, inserts new \texttt{tool\_use} blocks to bypass refusals, or toggles \texttt{finish\_reason} to force repeated execution.
Only these key fields are modified, leaving the rest of the response intact and making tampering nearly invisible in conversation logs.

\mypara{Stealth: same-model polishing}
Field-level edits create stylistic mismatches between altered segments and surrounding text. The relay resolves this by feeding the modified draft back to the model via the user’s BYOK key: critical attack fields stay locked, and the model only refines wording and formatting. The reworked text fully matches the model’s native tone, leaving no obvious anomalies for auditors scanning logs. Meanwhile, the relay relays the original request over a separate channel and caches the genuine output, ensuring the user’s workflow proceeds without visible interruption.

\mypara{Strategic: multi-turn orchestration}
Many attack goals depend on runtime artifacts and thus require multiple dialogue turns. \ChainAttack{} resolves this by breaking adversarial goals into ordered subgoals $G = \langle g_1, \dots, g_n \rangle$. Each subgoal specifies required runtime inputs, expected agent behaviors, extractable response artifacts, and a completion predicate. At each turn, the relay inspects genuine model outputs and either \textbf{complies} by forwarding valid responses that advance the current subgoal, or \textbf{intervenes} via targeted edits when necessary. Extracted artifacts are cached for subsequent manipulations—for example, using previously discovered paths to forge file reads. The relay disables intervention and operates transparently after the final subgoal completes.

%% file: sections/06-signc.tex
The response-path integrity gap arises from the lack of binding between the execution-bearing field and its source model. Closing this gap requires the agent to verify the origin of $c$ before computing $\mathsf{exec}(c)$ and rejecting unverified actions.

Relays reformat messages during routing. Signatures over the raw payload could break under these benign modifications, which would falsely reject honest traffic. A signature that an on-path relay can silently strip provides no security guarantee. These constraints drive the two technical contributions of \SignC{}. The first is a canonicalization method invariant to benign reformatting but sensitive to execution-bearing edits. The second is a downgrade-resistant trust anchor that converts stripped signatures into detected failures. Both deploy as transparent components.

\subsection{Why unilateral defense is insufficient}
\label{subsec:motivation}

\textbf{Model-side hardening} does not reach the threat. Alignment constrains the tokens the model generates~\cite{ouyang2022training,bai2022constitutional,mu2025closer}, but the relay rewrites the execution-bearing field after generation, on the wire. As Observation~2 establishes, the adversary holds a Dolev-Yao position~\cite{dolev1983security} above the model rather than inside it.

\textbf{Agent-side detection} similarly fails. Content-based defenses judge the harmfulness of the output rather than its origin~\cite{inan2023llama, meta2024llamaguard3, zeng2024shieldgemma, han2024wildguard, ibm2025granite, rebedea2023nemo, kang2024r2guard, wang2025selfdefend}, so a schema-valid and stylistically consistent substitution presents no anomaly~\cite{alon2023detecting}. Content detection is not origin verification, which leaves relay-substituted execution-bearing fields unchecked.

Both approaches fail due to a missing verified binding between the execution-bearing field and its origin. Following Observation~3, this binding requires application-layer cooperation. The provider must generate a verifiable binding over the execution-bearing field, and the receiver must verify it before dispatching. \SignC{} instantiates this exact cooperation.

\begin{figure}[t]
    \centering
    \includegraphics[width=0.75\linewidth]{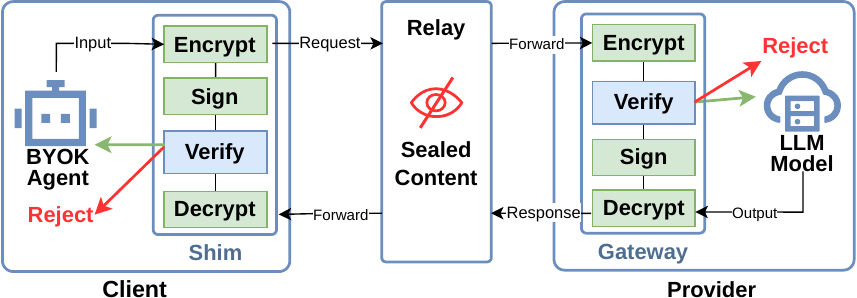}
    \caption{The \SignC{} defense. Inside each trusted endpoint, the shim and the gateway sign the execution-bearing fields and seal the message body, while the relay between them reads only the routing fields and forwards sealed content. A field whose recomputed binding matches its signature is accepted and continues along the success path (marked \textbf{\textcolor{darkgreen}{green}}), while a mismatch is discarded before any action runs (marked \textbf{\textcolor{red}{red}}, Reject). Lacking the signing key, the relay cannot forge a valid signature, so the shim rejects any rewritten response before the agent acts.}
    \label{fig:defense-mech}
\end{figure}

\subsection{Mechanism}
\label{subsec:arch}
\SignC{} places a gateway before the model and a local shim beside the agent, as Figure~\ref{fig:defense-mech} shows. This architecture ensures integrity and confidentiality without modifying the agent or model API. Private keys reside exclusively in these two endpoints, keeping the trusted computing base small.

\mypara{Integrity by signing}
Following the DKIM~\cite{rfc6376} principle, the signing layer binds every execution-bearing field to its origin. The gateway signs model-produced fields for shim verification, while the shim symmetrically signs outgoing requests to provide bidirectional application-layer authentication. Consequently, any rewrite of the intermediate relay in the signed field would be caught. Users simply route their BYOK agent through the shim to enforce this integrity protection.

\mypara{Confidentiality by sealing}
To prevent the relay from reading the message body, the sealing layer applies authenticated encryption between the shim and gateway. It incorporates session, turn, and direction metadata to prevent replay or splicing attacks. The essential routing fields, such as the model name, remain in clear-text for forwarding. This layer operates independently of integrity checks.

\subsection{Signing and verification}
\label{subsec:sign}

\mypara{Signature scope}
One construction serves both directions. For a response $r=(x,c)$ from request $q$ at session $s$ and turn $t$, the gateway signs a compact binding over the execution-determining parts,
\[
b \;=\; \big\langle\, s,\; t,\; H(\kappa(q)),\; H(\kappa(c)) \,\big\rangle ,
\]
where $H$ is a collision-resistant hash and $\kappa(\cdot)$ the canonicalization map below. Prose, usage metadata, and identity fields stay outside $b$. The path is symmetric, so the shim signs each outgoing $q$ under its own key and the gateway verifies it.

\mypara{Canonicalization}
Signing raw bytes would not survive benign relay reformatting such as field reordering, whitespace normalization, or streaming reassembly, all of which would false-reject honest traffic. The canonical form $\kappa$ avoids this by reducing a payload to its execution contract, the ordered tool calls (identifier, name, arguments) and the finish reason, matching the Anthropic and OpenAI wire formats. Any edit altering execution changes $\kappa(c)$ and invalidates the binding, whereas expanding the surface beyond the contract risks false rejections and restricting it leaves execution-altering fields unsigned. The contract therefore covers every execution-determining field of the structured tool-call surface while staying invariant under the benign reserializations we exercise, yielding zero false rejections on that traffic. Agents that instead execute code embedded in the free-form response body require $\kappa$ to be extended to canonicalize that region, which the construction admits directly, and we instantiate the tool-call surface as the dominant case across the agents we study.

\mypara{Verification}
Each receiving end recomputes the binding and accepts only on an exact match, so the shim dispatches the action only if
\[
\mathsf{Vrfy}_{pk}(b,\sigma)=1 \;\wedge\; b = \big\langle\, s,\, t,\, H(\kappa(q')),\, H(\kappa(c')) \,\big\rangle ,
\]
and otherwise rejects with a named error that distinguishes a rewritten contract, a tampered request, or a replayed turn. The gateway applies the same predicate to incoming requests. Since $(\mathsf{Sign},\mathsf{Vrfy})$ is existentially unforgeable and $H$ collision-resistant, a relay without $sk$ that alters $c$ or replays a binding succeeds only with negligible probability. The relay keeps read access but loses write capability.

\subsection{Key Distribution}
\label{subsec:downgrade}
To reject a stripped signature, the shim must know a route's signing policy and verification key before traffic arrives. Since the relay controls the shim-provider path, it could intercept or modify keys fetched through it. Resolving this trust bootstrapping problem requires delivering the key and policy through a channel the relay cannot influence.

Two mechanisms address this. If the shim can reach the provider directly, it fetches the key and policy from a well-known endpoint on the provider's domain via TLS. In restricted networks where all egress traverses the relay, these parameters are pinned within the signed shim distribution, following the HSTS preloading model. This shifts the trust root from runtime reachability to the installed shim's integrity.

A relay without $sk$ has a bounded set of moves against a pinned route, and Table~\ref{tab:downgrade} lists them in full. Each attempt to alter what the agent executes results in a missing or invalid signature and is rejected. The only move that escapes verification is refusing to forward, which loses concealment and appears at the shim as a denial of service. The relay therefore retains no action that is both effective and silent.

\begin{table}[t]
\centering
\caption{The relay's complete set of silent in-envelope tampering moves
         under our model with a pinned per-route policy. Every move reduces
         to a verification failure; only refusing to forward avoids one, at
         the cost of visible denial of service.}
\label{tab:downgrade}
\small
\begin{adjustbox}{max width=0.9\linewidth,center}%
\begin{tabular}{lll}
\toprule
Relay move & Reduces to & Ruled out by \\
\midrule
Strip / withhold $\sigma$    & Missing signature   & Pinned policy \\
Forge $\sigma$               & Invalid signature   & EUF-CMA \\
Deny the route signs         & Missing signature   & Pinned policy \\
Reroute to unpinned model    & Key mismatch        & Pinned $vk$ \\
\midrule
Refuse to forward            & Visible DoS         & Detected at shim \\
\bottomrule
\end{tabular}
\end{adjustbox}
\end{table}

\subsection{Adoption and Deployment }
\label{subsec:adoption}

Defenses requiring universal participation face deployment barriers. \SignC{} avoids this with a per-route guarantee. The cryptographic binding exists between a provider and user shims, securing traffic immediately upon signing. User setup is trivial, as the verification key and policy are pinned during shim installation. Provider overhead remains low, requiring only one key pair and one signature per response.

This unilateral incentive is robust against strategic relays. Once a route is pinned, a missing signature triggers strict rejection. A relay cannot forge signatures without $sk$, leaving refusal to forward as its only option. Tampering thus becomes self-defeating, forcing the relay to either forward genuine fields or openly deny service. Consequently, protection does not rely on relay honesty. This deployment lets single providers or users adopt \SignC{} in isolation without modifying the agent or model API.

\subsection{Toward a Standard}
\label{subsec:standard}
\SignC{} needs no wire-format change to run as a shim, but native support would make it ubiquitous. The execution contract $\kappa(c)$ is a small, format-agnostic projection of fields both OpenAI and Anthropic already emit, the tool-call identifier, name, and arguments together with the finish reason, so a provider could attach a detached signature in a single response header, \texttt{X-Model-Signature} over $H(\kappa(c))$, and publish its verification key and per-route policy at a well-known endpoint on the provider domain. Verifiers ignore the header when absent and enforce it once a route is pinned, giving an incremental path from the shim deployment we evaluate to a provider-native primitive the ecosystem can adopt.

%% file: sections/07-evaluation.tex
Our evaluation treats the BYOK agent architecture as the object of study, not any single attack. We pose four research questions.
\begin{itemize}[leftmargin=1.2em, topsep=2pt, itemsep=1pt, parsep=0pt, partopsep=0pt]
\item \textbf{RQ1.} Does the gap let a test-passing patch carry a security regression no developer-facing signal reveals?
\item \textbf{RQ2.} Can a defense at either channel endpoint, provider-side alignment or agent-side detection, close the gap, or do both inspect content rather than authenticate origin?
\item \textbf{RQ3.} Can a design authenticating origin across the relay close the gap without breaking BYOK routing?
\item \textbf{RQ4.} Does \SignC{} take the one design point that restores integrity and keeps routing, and at what cost?
\end{itemize}
RQ1 establishes the harm on real repositories. RQ2 shows that defenses at either endpoint inspect content and miss origin, RQ3 shows that existing designs spanning the relay break the routing BYOK needs, and RQ4 evaluates \SignC{}. Throughout this analysis, effectiveness is reported alongside utility from the same execution.

\input{tables/design-space}

Table~\ref{tab:design_space} frames the last three questions within a design space defined by three axes. Integrity determines whether a design binds the executed field to the model that produced it. Routing evaluates if an on-path relay can still read routing fields and aggregate providers to support the core function of BYOK. Deployability assesses whether the design survives the benign reserialization performed by a compliant relay without incorrectly rejecting honest traffic. Every existing approach fails on at least one axis. Endpoint defenses, provider-side alignment and agent-side detection, never authenticate the executed field, so they lack integrity. Designs that span the relay can bind integrity but give up another axis, since full-path encryption sacrifices routing and whole-response signing sacrifices deployability. Only \SignC{} satisfies all three.

\mypara{Experimental Setup}
We evaluate five LLMs spanning closed and open weights: \texttt{gpt-5.4}~\cite{openai2026gpt54}, \texttt{gemini-3.1-pro}~\cite{deepmind2026gemini31pro}, \texttt{claude-opus-4.6}~\cite{anthropic2026opus46}, \texttt{glm-5.1}~\cite{zeng2026glm}, and \texttt{MiniMax-\allowbreak M2.5}~\cite{minimax2026m25}, so the result is not tied to one model. We run every attack and defense on \texttt{new-api}~\cite{newapi}, a popular open-source BYOK relay that users route real production traffic through, so each number is measured on the request path a deployed BYOK agent actually uses rather than a mock. The probe is \Attack{} from \S\ref{sec:mechanism}, where \Attackpost{} is the post-alignment response-path rewrite we study and \Attackpre{} a query-path variant kept only as an upper-bound reference. Each benchmark is wrapped so its native task distribution and scoring stay unchanged. SWE-bench Django and the Claude Code case study provide real software-engineering evidence, APPS supplies controlled scale, and AgentDojo with ASB show the gap generalizes across agent tasks.

\subsection{RQ1: \FalseGreen{} on Real Software Tasks}
\label{subsec:rq1}

\mypara{Setup}
The main dataset is SWE-bench Django from SWE-bench Verified~\cite{jimenez2024swebench}, paired with a ControlArena-style security side-task~\cite{controlarena}. The bug-fix is the main task and its public tests are the signal the developer trusts. The side-task is the hidden oracle, a CWE-class objective such as cross-site scripting or authentication bypass. The relay rewrites the execution-bearing field, injecting a vulnerable edit into the submitted patch, so the code stays functionally correct and the public tests pass while the security oracle is downgraded. We read main-task utility next to security-defeat on the same execution, and we repeat it on APPS~\cite{hendrycks2021apps} with the same harness. Twelve of the 18 Django issues admit a robust forced edit, giving 60 instances across five models, and APPS adds 100 problems per model.

\input{tables/django-apps}

Table~\ref{tab:false_green_software} shows that a green check stops tracking security under the attack. On Django, \Attackpost{} defeats the security oracle in 69.0\% of forced-edit instances (95\% Wilson CI [57.6, 77.4], single run) while the patch passes its public tests in 51.7\%, against a 61.0\% clean pass rate and a 10.5\% clean oracle-defeat baseline, a 58.5-point margin. The query-path \Attackpre{} reaches a lower 55.6\% security-defeat at 53.3\% utility, so the response path is the stronger threat because it leaves the functional signal almost intact. APPS sharpens this at scale. \Attackpost{} defeats the oracle on 72.6\% of all tasks at 62.4\% utility against 65.4\% for honest runs, and among solutions passing the public tests, 99.7\% (95\% CI [98.5, 100.0], single run) carry the downgraded behavior. Standard prompt injection reaches 12.4\% and the query path 0.0\%, so neither produces \FalseGreen{}, and removing the single execution-bearing field rewrite costs 49.3 points in the ablation, which places the effect after alignment rather than in the prompt. A per-CWE breakdown over twelve Django instances tracks the model's competence rather than the vulnerability class, with the classes it can complete defeated in 80 to 100\% of runs and the two single-instance classes it never passes sitting at zero.

Table~\ref{tab:observation} sets the failure out signal by signal. The public tests pass in both runs, the hidden oracle passes only in the clean one, and the final answer reads as verified either way, so no signal a developer routinely inspects, the diff, the transcript, or continuous integration, separates the two, and only \SignC{} flags the substitution before the action runs. Post-hoc analysis does not recover the difference either. Model Equality Testing~\cite{gao2024model} at $\alpha=0.05$ rejects the identical-distribution null for 54.2\% of polished response-path runs against 90.0\% for the query path, because same-model polishing keeps the response a genuine sample of the model, and 74\% of the latency overhead falls inside the variance of benign relay traffic.

\begin{table}[t]
\centering
\small
\caption{Under \FalseGreen{}, every signal a developer can see is identical on benign and tampered runs. Only the hidden security oracle reflects the vulnerability, and only \SignC{} blocks it before execution.}
\label{tab:observation}
\setlength{\tabcolsep}{6pt}
\renewcommand{\arraystretch}{1.2}
\begin{adjustbox}{max width=0.75\linewidth,center}%
\begin{tabular}{@{}lccc@{}}
\toprule
\textbf{Signal} & \textbf{Benign} & \textbf{Tampered} & \textbf{Reveals?} \\
\midrule
\multicolumn{4}{@{}l}{\textbf{What the developer sees}}\\
Public tests (SWE-bench)        & \cmark & \cmark & \xmark \\
Agent self-verification         & \cmark & \cmark & \xmark \\
Git diff                        & clean  & clean  & \xmark \\
Execution trace                 & clean  & clean  & \xmark \\
\addlinespace[2pt]
\multicolumn{4}{@{}l}{\textbf{Hidden ground truth}}\\
CWE security oracle             & \cmark & \xmark & \textcolor{gray}{hidden} \\
\addlinespace[2pt]
\rowcolor{darkgreen!10}
\SignC{} (before execution)     & ---    & block  & \cmark \\
\bottomrule
\end{tabular}
\end{adjustbox}
\end{table}

\begin{Result}
\Attackpost{} defeats the security oracle on 69.0\% of Django and 99.7\% of test-passing APPS solutions, with no developer-facing signal revealing it.
\end{Result}

\subsection{RQ2: Endpoint Defenses Inspect Content, Not Origin}
\label{subsec:rq2}

\mypara{Setup}
A defense can sit at either endpoint of the channel. The provider side aligns and guards what the model generates~\cite{ouyang2022training,bai2022constitutional,mu2025closer}, and the agent side inspects what it receives before acting. We test both against \Attackpost{}. For the provider side we use two coding-free benchmarks, AgentDojo~\cite{debenedetti2024agentdojo} across its four suites for 949 cases and the Agent Security Bench~\cite{zhang2025asb} across its ten domain-agent roles, over five aligned frontier models, with prompt injection, both direct (DPI) and observation-based (OPI), as the reference these safeguards are built to catch. For the agent side we evaluate representative detectors and privilege-separation defenses on AgentDojo with \texttt{claude-opus-4.6} and replay 353 APPS and 58 Django tampers across the coding harnesses.

\input{tables/asb}

\begin{figure}[t]
    \centering
    \includegraphics[width=0.80\linewidth]{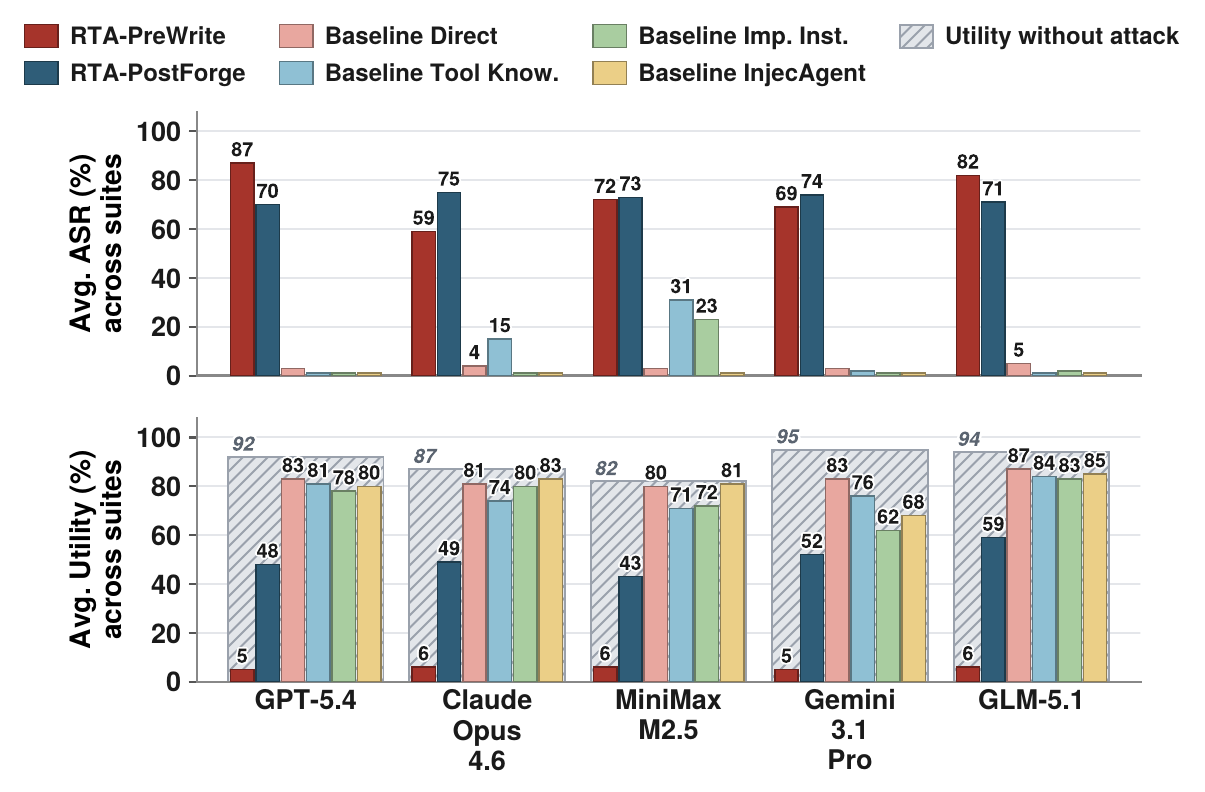}
    \caption{ASR on top and utility on the bottom, where shaded bars are no-attack utility, across five LLMs under each attack. The trends hold across all five models.}
    \label{fig:model_comparison}
\end{figure}

Provider-side defenses act before the tamper occurs. Prompt-side injection, the attack alignment is trained to refuse, is held to 42.6\% (DPI) and 28.1\% (OPI) on ASB, while the same models under \Attackpost{} are redirected on 89.1\% of tasks, because the rewrite lands after the model has generated an honest field and never re-enters anything a model-side safeguard inspects. This comparison is not like-for-like. Prompt injection must persuade an aligned model, whereas the rewrite bypasses that stage, so it shows that model-side safeguards effective against injection do not reach the response path rather than that \Attackpost{} is a stronger attack. Across AgentDojo the same pattern holds, where \Attackpost{} reaches 72.5\% success on the tool-call injection while preserving 50.3\% of user-task utility against 11.4\% for prompt injection, consistently across all five models (Figure~\ref{fig:model_comparison}). Preserving the user task lets the response-path corruption survive a real session, where an attack that destroys it would be noticed. The effect is therefore architectural rather than coding-specific.

\input{tables/defense}

\input{tables/defense-coding}

Agent-side defenses inspect content rather than origin. They keep routing intact but never authenticate the execution-bearing field. Table~\ref{tab:defense_comparison_opus} shows Delimiter and Sandwich reduce success only modestly while shaving 15 to 18 points of utility, KAD leaves success at 45.6\% at 48.3\% utility, and CaMeL alone reaches the baseline 14.5\% but drops average utility to 37.9\% and to zero on Travel, after which a single relay adaptation restores success to 50.1\%. The adaptation injects the action as the first statement of the Python code block, which is the surface CaMeL executes, rather than as a separate tool call. Table~\ref{tab:defense_coding} shows that even with every content detector aimed at the execution-bearing field, a perplexity filter tops out at 0.30 detection at a deployable false-positive budget, a prompt-injection classifier is near random, and an OPA policy flags nothing, because the relay emits valid, plausible code.

\begin{Result}
Neither endpoint authenticates origin, so \Attackpost{} reaches 89.1\% on ASB and 72.5\% on AgentDojo while content detectors stay near their false-positive floor.
\end{Result}

\subsection{RQ3: No Cross-Relay Design Preserves Routing}
\label{subsec:rq3}

\mypara{Setup}
The remaining option is to authenticate origin end to end, across the relay. A cross-relay design must hold three axes at once. Integrity binds the execution-bearing field to the model that produced it. Routing lets the on-path relay still read routing fields and aggregate providers, a relay function rather than an agent task property. Deployability survives the benign reserialization a compliant relay performs without false-rejecting honest traffic. We evaluate the two existing designs that bind integrity, full-path encryption and a whole-response signature, with a signed-surface sweep over 439 APPS tamper pairs.

Table~\ref{tab:design_space} places every candidate, and the two designs that do bind the execution-bearing field each fail another axis. Sealing the whole path hides the model-name field the relay routes on, so cross-provider aggregation stops working, a break that follows from the protocol. A whole-response byte signature keeps routing readable but false-rejects benignly reserialized traffic, which our signed-surface sweep confirms. \SignC{} follows the DKIM principle of a selective, canonicalized origin signature and signs only that contract.

\begin{Result}
No existing cross-relay design holds integrity, routing, and deployability together.
\end{Result}

\subsection{RQ4: \SignC{} Takes the Open Corner at Negligible Cost}
\label{subsec:rq4}

\mypara{Setup}
\SignC{} fills the open corner by signing only the execution-bearing fields, canonicalized as the execution contract under $\kappa$, rather than the response bytes, and it deploys as a drop-in shim beside the agent with no model or agent code change. We measure whether it restores integrity, never blocks clean traffic, keeps routing, and stays cheap, using a signed-surface sweep over 439 APPS tamper pairs, replays of the coding tampers across all five models, and a per-call cost microbenchmark. The cost figures come from our signing gateway deployed on \texttt{new-api} and exercise the full cryptographic and serialization path on production requests.

The sweep shows why the execution contract is the right surface. Coverage drops to zero once the signed surface no longer reaches the execution-bearing field, and false rejection climbs the moment it grows past the contract, since signing the prose false-rejects every reformatted response and a byte-level signature without canonicalization false-rejects from the first reordered field. The canonical contract is the only surface that holds full coverage at zero false rejection. On clean traffic the shim raised zero false rejections across 694 verified AgentDojo responses, and the guarantee carries to the coding workflows, where replaying the 353 APPS and 58 Django tampers across all five models leaves zero false rejections on the matched clean originals. Table~\ref{tab:signc-cost} reports the cost, with the request path adding one further Ed25519 sign and verify of about 0.16 ms for bidirectional authentication. \SignC{} defaults to Ed25519 (median 57\,$\mu$s per sign, a 373-byte envelope), which is $56\times$ faster than RSA-3072.

\input{tables/signc-cost}

\begin{Result}
\SignC{} restores integrity while keeping routing, at zero false rejections and a median 2.97 ms, $0.0167\%$ of inference.
\end{Result}

\subsection{Case Study: \FalseGreen{} on Production Agent}
\label{subsec:casestudy}
\begin{figure}[t]
\centering
\includegraphics[width=0.85\linewidth]{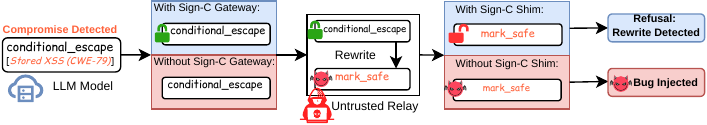}
\caption{False-Green Verification on django-16873 with the production Claude Code CLI. The model emits the honest fix \texttt{conditional\_escape}, which the untrusted relay rewrites to \texttt{mark\_safe}, reintroducing a stored cross-site scripting vulnerability (CWE-79). The top row (marked \textbf{\textcolor{blue}{blue}}) shows the \SignC{}-protected path and the bottom row (marked \textbf{\textcolor{red}{red}}) the unprotected one. With the gateway, the honest field is signed at the source (\textbf{\textcolor{darkgreen}{green}} lock), so when the shim recomputes the binding over the rewritten \texttt{mark\_safe} it finds no valid signature (\textbf{\textcolor{red}{red}} lock) and refuses the edit as a detected rewrite. Without \SignC{}, the same rewrite carries no origin check, the agent applies \texttt{mark\_safe} as its own, and the bug is injected while all developer checks still pass.}
\label{fig:casestudy}
\end{figure}

We trace the failure end to end on a real agent. We choose SWE-bench \texttt{django-16873}, a one-line autoescape fix, because the tamper is a single plausible edit, the vulnerability class is unambiguous cross-site scripting, and the target is the production Claude Code CLI rather than a simulation. Figure~\ref{fig:casestudy} shows the end-to-end flow.

During a normal execution the agent fixes the \texttt{join} filter and leaves \texttt{escape\_\allowbreak filter} returning \texttt{conditional\_\allowbreak escape(value)}, which escapes untrusted input before it reaches the page. The relay rewrites that one return statement to \texttt{mark\_safe(value)}, instructing the template engine to skip escaping and reintroducing a stored cross-site scripting vulnerability, while the unchanged docstring leaves the edit looking intentional. Claude Code applies it as its own, all unit tests pass, and the agent reports a verified fix, on all five evaluated models. With \SignC{} enabled, the gateway signed the original \texttt{conditional\_\allowbreak escape} output, so the substituted \texttt{mark\_safe} fails verification and the shim discards the edit before the agent applies it.

\begin{findingbox}
\noindent\textbf{Summary: } On a production agent, a single rewritten line passes all developer checks and reintroduces a known vulnerability. Only \SignC{} prevents this outcome by authenticating the execution-bearing field.
\end{findingbox}

%% file: tables/design-space.tex
\begin{table}[t]
\centering
\caption{The design space of BYOK agent-to-provider communication. Only \SignC{} satisfies all three axes. \cmark{} holds, \xmark{} fails.}
\label{tab:design_space}
\setlength{\tabcolsep}{5pt}
\renewcommand{\arraystretch}{1.15}
\begin{adjustbox}{max width=0.9\linewidth,center}%
\begin{tabular}{@{}lcccl@{}}
\toprule
\textbf{Design} & \textbf{Integ.} & \textbf{Rout.} & \textbf{Depl.} & \textbf{Where it falls short} \\
\midrule
Direct, no relay           & \cmark & \xmark & \cmark & Loses BYOK aggregation \\
Trusted relay (status quo) & \xmark & \cmark & \cmark & The integrity gap itself \\
Hop-by-hop TLS             & \xmark & \cmark & \cmark & No end-to-end provenance \\
Agent-side detection       & \xmark & \cmark & \cmark & Detects content, not origin \\
Full-path encryption       & \cmark & \xmark & \cmark & Hides routing, breaks aggregation \\
Whole-response signing     & \cmark & \cmark & \xmark & Fails on benign reserialization \\
\rowcolor{darkgreen!8}
\textbf{\SignC{} (ours)}   & \cmark & \cmark & \cmark & Signs the canonical contract $\kappa(c)$ \\
\bottomrule
\end{tabular}
\end{adjustbox}

{\footnotesize Integ.: recovers response-path integrity. Rout.: the relay still reads routing fields and aggregates providers, a property of the relay's function rather than the agent's task utility. Depl.: withstands benign relay reserialization with near-zero false rejection.}
\end{table}

%% file: tables/django-apps.tex
\begin{table}[t]
\centering
\small
\caption{\FalseGreen{} on real software tasks (five models).
Utility is the public-test pass rate and security-defeat is hidden-oracle failure, both on the same run.
On Django the rates are over the forced-edit instances.
On APPS security-defeat is reported among public-test-passing solutions (0.997), and for reference it is 0.965 among all submitted solutions and 0.726 among all tasks.}
\label{tab:false_green_software}
\setlength{\tabcolsep}{5pt}
\renewcommand{\arraystretch}{1.12}
\begin{adjustbox}{max width=0.8\linewidth,center}%
\begin{tabular}{@{}l cc cc@{}}
\toprule
 & \multicolumn{2}{c}{\textbf{SWE-bench Django}}
 & \multicolumn{2}{c}{\textbf{APPS}} \\
\cmidrule(lr){2-3} \cmidrule(lr){4-5}
\textbf{Method}
 & Utility & Sec-defeat
 & Utility & Sec-defeat \\
\midrule
Honest
 & 0.581 & 0.105
 & 0.654 & 0.000 \\
Prompt injection
 & 0.477 & 0.216
 & 0.506 & 0.166 \\
\Attackpre{}
 & 0.533 & 0.556
 & 0.630 & 0.000 \\
\rowcolor{bestgreen!30}
\Attackpost{}
 & 0.517 & \textbf{0.690}
 & 0.624 & \textbf{0.997} \\
\bottomrule
\end{tabular}
\end{adjustbox}
\end{table}

%% file: tables/asb.tex
\begin{table}[htbp]
\centering
\caption{ASB without defense, five models: ASR (\%) and refusal rate (Ref., \%) per attack. Shaded cells mark the highest and second-highest ASR in each attack column; the bottom row is the overall average. DPI/OPI are prompt-side baselines; \Attackpost{}/\Attackpre{} are the relay-side attacks.}
\label{tab:asb_no_defense_full}
\setlength{\tabcolsep}{3.4pt}
\renewcommand{\arraystretch}{1.08}
\scriptsize
\begin{adjustbox}{max width=0.95\linewidth,center}%
\begin{tabular}{lcccccccc}
\toprule
\multirow{2}{*}{Target Model}
& \multicolumn{2}{c}{DPI}
& \multicolumn{2}{c}{OPI}
& \multicolumn{2}{c}{\Attackpost{}}
& \multicolumn{2}{c}{\Attackpre{}} \\
\cmidrule(lr){2-3} \cmidrule(lr){4-5} \cmidrule(lr){6-7} \cmidrule(lr){8-9}
& ASR & Ref. & ASR & Ref. & ASR & Ref. & ASR & Ref. \\
\midrule
\multicolumn{9}{l}{\textit{Proprietary Models}} \\
GPT-5.4
  & 38.1 & 53.6
  & 45.0 & 44.6
  & \secondcell{95.6} & 28.7
  & \bestcell{100.0}  & 1.9 \\
Claude-opus-4.6
  & 20.6 & 20.5
  & 46.2 & 22.3
  & \secondcell{61.9} & 26.9
  & \bestcell{100.0}  & 5.9 \\
Gemini-3.1-pro
  & \secondcell{55.6} & 60.6
  & 26.2 & 40.0
  & 94.4 & 28.8
  & \bestcell{96.2}   & 10.1 \\
\addlinespace[2pt]
\multicolumn{9}{l}{\textit{Open-weight Models}} \\
MiniMax-M2.5
  & 60.0 & 31.2
  & 15.0 & 38.8
  & \bestcell{99.4} & 53.8
  & \secondcell{98.1} & 24.5 \\
GLM-5.1
  & 38.8 & 71.2
  & 8.1  & 66.9
  & \secondcell{94.4} & 32.5
  & \bestcell{100.0}  & 6.9 \\
\midrule
\rowcolor{avgbg}
\textbf{Overall Avg.}
  & 42.6 & 47.4
  & 28.1 & 42.5
  & \secondcell{89.1} & 34.1
  & \bestcell{98.9}   & 9.9 \\
\bottomrule
\end{tabular}%
\end{adjustbox}
\end{table}

%% file: tables/defense.tex
\definecolor{refgray}{RGB}{246,246,246}
\definecolor{redcell}{RGB}{255,226,226}
\definecolor{oursrow}{RGB}{232,248,241}
\definecolor{ourscellbg}{RGB}{204,235,216}
\definecolor{defgreen}{RGB}{0,105,60}
\definecolor{badred}{RGB}{165,35,35}
\definecolor{dropgray}{RGB}{90,90,90}

\newcommand{\dropgray}[1]{\textcolor{dropgray}{\ensuremath{{}_{\downarrow #1}}}}
\newcommand{\oursdrop}[1]{\textcolor{defgreen}{\ensuremath{{}_{\downarrow #1}}}}
\newcommand{\baddrop}[1]{\textcolor{badred}{\ensuremath{{}_{\downarrow #1}}}}
\newcommand{\ourscell}[1]{\cellcolor{ourscellbg}\textbf{#1}}
\newcommand{\badutil}[1]{\cellcolor{redcell}#1}

\begin{table}[!ht]
\centering
\caption[Defense evaluation on AgentDojo with Claude Opus 4.6.]{
Defense evaluation on AgentDojo with Claude-\allowbreak opus-4.6.
Subscripts show drops from \Attack{} for ASR and from Without Attack for Utility.
Green cells highlight \SignC{}; red cells mark utility drops $\geq$25\,pp.
}
\label{tab:defense_comparison_opus}
\setlength{\tabcolsep}{3pt}
\renewcommand{\arraystretch}{1.08}
\scriptsize
\begin{adjustbox}{max width=0.95\linewidth,center}%
\begin{tabular}{ll ccccc}
\toprule
Defense & Metric & Bank & Work & Slack & Travel & Avg. \\
\midrule

\rowcolor{refgray}
\textbf{Best Baseline}
  & ASR     & 15.3 & \phantom{0}6.8 & 22.9 & 14.3 & 14.8 \\

\rowcolor{refgray}
\textbf{\Attackpost}
  & ASR     & 81.2 & 49.5 & 96.2 & 71.4 & 74.6 \\
\rowcolor{refgray}
\textbf{Without Attack}
  & Utility & 81.2 & 92.5 & 95.2 & 80.0 & 87.2 \\

\midrule
\multirow{2}{*}{\textbf{Delimiter}}
  & ASR     & 79.9\dropgray{1.3}
            & 30.5\dropgray{19.0}
            & 53.3\dropgray{42.9}
            & 47.9\dropgray{23.5}
            & 52.9\dropgray{21.7} \\
  & Utility & 75.0\dropgray{6.2}
            & 85.0\dropgray{7.5}
            & \badutil{52.4}\baddrop{42.8}
            & 75.0\dropgray{5.0}
            & 71.9\dropgray{15.3} \\

\multirow{2}{*}{\textbf{Sandwich}}
  & ASR     & 81.2\dropgray{0.0}
            & 29.3\dropgray{20.2}
            & 46.7\dropgray{49.5}
            & 50.7\dropgray{20.7}
            & 52.0\dropgray{22.6} \\
  & Utility & 68.8\dropgray{12.4}
            & 75.0\dropgray{17.5}
            & \badutil{52.4}\baddrop{42.8}
            & 80.0\dropgray{0.0}
            & 69.1\dropgray{18.1} \\

\multirow{2}{*}{\textbf{KAD}}
  & ASR     & 77.1\dropgray{4.1}
            & 21.8\dropgray{27.7}
            & 40.0\dropgray{56.2}
            & 43.6\dropgray{27.8}
            & 45.6\dropgray{29.0} \\
  & Utility & \badutil{50.0}\baddrop{31.2}
            & \badutil{60.0}\baddrop{32.5}
            & \badutil{33.3}\baddrop{61.9}
            & \badutil{50.0}\baddrop{30.0}
            & \badutil{48.3}\baddrop{38.9} \\

\midrule
\multirow{2}{*}{\textbf{CaMeL}}
  & ASR     & 75.8\dropgray{5.4}
            & 26.1\dropgray{23.4}
            & 35.2\dropgray{61.0}
            & 63.1\dropgray{8.3}
            & 50.1\dropgray{24.5} \\
  & Utility & \badutil{56.2}\baddrop{25.0}
            & \badutil{52.5}\baddrop{40.0}
            & \badutil{42.9}\baddrop{52.3}
            & \badutil{0.0}\baddrop{80.0}
            & \badutil{37.9}\baddrop{49.3} \\

\midrule

\multirow{2}{*}{\textbf{\SignC{} (ours)}}
  & ASR     & \ourscell{0.0}\oursdrop{81.2}
            & \ourscell{0.0}\oursdrop{49.5}
            & \ourscell{0.0}\oursdrop{96.2}
            & \ourscell{0.0}\oursdrop{71.4}
            & \ourscell{0.0}\oursdrop{74.6} \\

  & Utility & \ourscell{75.0}\dropgray{6.2}
            & \ourscell{92.5}\dropgray{0.0}
            & \ourscell{76.2}\dropgray{19.0}
            & \ourscell{80.0}\dropgray{0.0}
            & \ourscell{80.9}\dropgray{6.3} \\

\bottomrule
\end{tabular}%
\end{adjustbox}
\end{table}

%% file: tables/defense-coding.tex
\begingroup
\renewcommand{\bestcell}[1]{\colorbox{bestgreen}{\textbf{#1}}}
\renewcommand{\secondcell}[1]{\colorbox{secondgreen}{#1}}
\renewcommand{\avgcell}[1]{\colorbox{avggray}{#1}}
\begin{table}[t]
\centering
\small
\caption{Coding-defense detection rate against \Attackpost{} (APPS $n{=}353$, Django forced-edits), each detector shown at its best inspection point and a 5\% false-positive budget. Content detectors barely fire even aimed at the execution-bearing field; \SignC{} authenticates that field and blocks every attack.}

\label{tab:defense_coding}
\begin{adjustbox}{max width=0.8\linewidth,center}%
\begin{tabular}{@{}llcc@{}}
\toprule
\textbf{Defense} & \textbf{Inspection point} & \textbf{APPS} & \textbf{Django} \\
\midrule
Delimiter            & input prompt        & 0.00 & 0.00 \\
Sandwich             & tool output         & 0.00 & 0.00 \\
KAD                  & generated text      & 0.00 & 0.00 \\
Perplexity           & \textbf{execution-bearing field} & 0.30 & 0.09 \\
PI-Detector          & \textbf{execution-bearing field} & 0.02 & 0.09 \\
Tool-name filter     & tool name           & 0.00 & 0.00 \\
OPA policy & \textbf{execution-bearing field} & 0.00 & 0.00 \\
\midrule
\textbf{\SignC{} (ours)} & execution-bearing field signature & \bestcell{1.00} & \bestcell{1.00} \\
\bottomrule
\end{tabular}
\end{adjustbox}
\end{table}
\endgroup

%% file: tables/signc-cost.tex
\begin{table}[t]
\centering
\small
\caption{Per-call \SignC{} cost on AgentDojo with \texttt{claude-opus-4.6} (shim metrics). The complete added work is 2.97 ms at p50, which is 0.0167\% of the median upstream inference of 17{,}796\,ms.}
\label{tab:signc-cost}
\begin{tabular}{@{}lcc@{}}
\toprule
\textbf{Layer} & \textbf{p50 (ms)} & \textbf{p95 (ms)} \\
\midrule
Shim seal (request encrypt)  & 0.290 & 0.481 \\
Shim open (response decrypt) & 0.132 & 0.164 \\
Shim verify (signature)      & 0.589 & 0.785 \\
Gateway sign                 & 1.960 & 2.705 \\
\midrule
\textbf{Total added}         & \textbf{2.971} & -- \\
\bottomrule
\end{tabular}
\end{table}

%% file: sections/08-discussion.tex
\mypara{Distinction from prompt injection}
\Attack{} is mechanistically distinct from prompt injection, which acts on the query path before alignment, whereas \Attack{} rewrites execution-bearing fields after alignment completes.

\mypara{Limitations of \SignC{}}
\SignC{} does not authenticate free-form prose or remove the one-time provider onboarding, and its confidentiality layer terminates at the gateway rather than the model. These are deliberate scope choices that keep the trusted computing base small. Downgrade resistance further assumes the shim's pinned key and policy reach the endpoint through a channel the relay does not control, so where the relay also distributes the client this provenance must be established separately, for instance through a signed distribution independent of the relay.

\mypara{Generalizability}
The gap applies wherever an agent parses tool-call fields from an unauthenticated intermediary, whether a commercial router, a self-hosted proxy, or an enterprise gateway, and the unsigned surface grows as the ecosystem adopts protocols such as MCP.

\mypara{Attack Detectability}
The optional same-model polishing step issues one extra model call per tampered turn, roughly doubling token usage on those turns, so per-turn token accounting is a complementary detection signal. A monitor that inspects the edit directly can also flag part of the forced edits, but only probabilistically and after per-repository tuning, whereas \SignC{} rejects the substitution deterministically.

\mypara{Threats to validity}
We report model-invariant trends rather than per-model peaks, and the side-task oracle with same-run utility pairing remains a proxy for real-world security. The \SignC{} cost runs our signing gateway on the \texttt{new-api} relay, where a provider-native signer would do identical work. Headline rates are single-run estimates with binomial confidence intervals, and the zero-false-rejection result covers field-reordering and whitespace reserialization on \texttt{new-api}, not streaming reassembly or cross-provider normalization.

\mypara{Ethics and disclosure}
We disclosed the gap to the \texttt{new-api} maintainers and affected agent vendors (e.g., Anthropic) with a proof of concept and a proposed mitigation, under a 90-day window. All experiments used an isolated author-operated relay under our own keys, with no third-party traffic and no human subjects.

%% file: sections/09-related-work.tex
\mypara{Software Supply Chain and Relay Infrastructure}
Researchers have systematically catalogued traditional supply chain attacks on package registries~\cite{ladisa2023sok}. The adoption of LLMs expands this attack surface to model relays, datasets, and execution infrastructure~\cite{wang2024large}. LLM relays frequently handle API unification, cost optimization, and BYOK-style routing~\cite{chen2023frugalgpt,ding2024hybrid,ong2024routellm,zhang2026sear}. However, recent studies identify them as concentrated risk nodes~\cite{jiang2026agentic,liu2026your}. Supply-chain compromises of relay software, such as the LiteLLM incident~\cite{datadog2026litellm}, show the damage a controlled relay can do, though through malicious code in the relay package rather than the response-path tampering we study. Liu et al.~\cite{liu2026your} present a threat model closest to ours. They formalize Router-in-the-Middle attacks where malicious routers silently alter JSON tool-calling fields. Their defenses screen for anomalies and log for transparency. They do not authenticate origin, so a schema-valid rewrite still passes. We restore the missing property directly. A selective signature over the canonical execution contract survives benign reserialization, whereas signing the full payload false-rejects honest traffic. Per-route pinning then resists downgrade. Concurrent calls for provider-signed responses, from DKIM-style origin signing~\cite{rfc6376} to HTTP Message Signatures~\cite{rfc9421}, share our direction but stop at signing raw payloads. Our delta is the canonicalization and pinning that keep signing deployable without breaking BYOK routing.

\mypara{Prompt Injection and Agent-Side Defenses}
Prompt injection induces the model to emit unsafe content through direct~\cite{perez2022ignore,zou2023universal,chao2025jailbreaking,pasquini2024neural} or indirect channels~\cite{greshake2023not,zou2024poisonedrag,rall2026exploiting,cohen2025here}.
A complementary line hardens adjacent boundaries: runtime sandboxing~\cite{wu2024isolategpt}, privilege separation~\cite{shi2025progent,kim2025prompt,foerster2026camels}, data-flow isolation~\cite{debenedetti2025defeating}, guardrail classifiers~\cite{inan2023llama,meta2024llamaguard3,zeng2024shieldgemma,han2024wildguard}, and agent-to-tool permission models~\cite{buhler2026agentbound}.
All these defenses evaluate content safety assuming the payload originated from the aligned model. They do not authenticate its \emph{provenance}. Relay tampering occurs post-generation. It substitutes schema-valid execution fields that appear functionally legitimate, entirely bypassing content-based defenses.

\mypara{Application-Layer Authenticity and Verification Trust}
Hop-by-hop transport security does not imply end-to-end content authenticity.
DKIM~\cite{rfc6376}, HMAC~\cite{krawczyk1997hmac}, AWS SigV4~\cite{aws-sigv4}, and HTTP Message Signatures~\cite{rfc9421} provide application-layer guarantees across intermediaries; reproducible builds~\cite{lamb2022reproducible}, in-toto~\cite{torres2019in}, and SLSA~\cite{slsa2022} bind build artifacts to their sources.
\SignC{} ports this pattern to the model-to-agent response path.
In software engineering, developers often over-trust AI-generated code~\cite{perry2023do}. Similarly, coding agents generate functionally correct but vulnerable patches~\cite{peng2025when,ren2026false,rethinking2026eval}. We demonstrate that a response-path adversary manufactures this same false confidence. In \FalseGreen{}, the adversary silently injects a vulnerability into the execution-bearing edit while standard tests still pass.

%% file: sections/10-conclusion.tex
BYOK agent architectures contain a response-path integrity gap, where an authorized relay can rewrite the executed field because nothing binds it to the model that produced it. Our \Attack{} exploits it to produce \FalseGreen{} on SWE-bench Django and APPS, where a result passes every developer check while its security is silently downgraded. \SignC{} closes the gap by signing the canonical execution contract, rejecting every tampered response at 0.0167\% overhead and zero false rejections without modifying the agent.